Miscibility of blends of poly(methyl methacrylate) and oligodiols based on a bisphenol A nucleus and ethylene oxide or propylene oxide branches

B. Jaffrennou[a], E. R. Soulé[b], F. Méchin[a], J. Borrajo[b], J. P. Pascault[a],

R. J. J. Williams[b,*]

[a] *Institut National des Sciences Appliqués de Lyon (INSA), Laboratoire des Matériaux Macromoléculaires/IMP, UMR CNRS Nº 5627, Bât. Jules Verne, 20 Av. A. Einstein, 69621 Villeurbanne Cedex, France*

[b] *Institute of Materials Science and Technology (INTEMA), University of Mar del Plata and National Research Council (CONICET), Av. J. B. Justo 4302, 7600 Mar del Plata, Argentina*

**Abstract**

Cloud-point curves of blends of poly(methyl methacrylate) (PMMA) with a series of oligodiols based on a bisphenol A nucleus and short branches of poly(ethylene oxide) or poly(propylene oxide) (BPA-EO or BPA-PO), and with PEO and PPO oligomers, were obtained using a light transmission device. Experimental results were fitted with the Flory-Huggins model using an interaction parameter depending on both temperature and composition. For PMMA/PEO and PMMA/PPO blends, the miscibility increased when increasing the size of the diol, due to the significant decrease in the entropic and enthalpic terms contributing to the interaction parameter. This reflected the decrease in the self-association of solvent molecules and in the contribution of terminal OH groups to the mismatching of solubility parameters. For PMMA/BPA-EO blends, a decrease of the entropic contribution to the interaction parameter when increasing the size of the oligodiol was also found. However, the effect was counterbalanced by the opposite contribution of combinatorial terms leading to cloud-point curves located in approximately the same temperature range. For PMMA/BPA-PO blends, the interaction parameter exhibited a very low value. In this case, the effect of solvent size was much more important on combinatorial terms than on the interaction parameter, leading to an increase in miscibility when decreasing the oligodiol size. For short BPA-PO oligodiols no phase separation was observed. The entropic contribution of the interaction parameter exhibited an inverse relationship with the size of the oligodiols, independent of the nature of the chains bearing the hydroxyls and the type of OH groups (primary or secondary). This indicates that the degree of self-association of solvent molecules through their OH terminal groups, was mainly determined by their relative sizes.

# 1. Introduction

Blends of poly(methyl methacrylate) (PMMA) in reactive solvents based on diglycidylether of bisphenol A (DGEBA) and an appropriate co-monomer, have been the subject of several recent studies [1-8]. PMMA and the DGEBA/co-monomer form homogeneous solutions in the whole composition range, that in most cases phase-separate in the course of polymerization.

In order to extend these studies to other chemistries, different types of reactive solvents for PMMA were tested. Here, experimental results obtained for the solubility of PMMA in a series of oligodiols based on a bisphenol A nucleus and short branches of poly(ethylene oxide) or poly(propylene oxide) (BPA-EO or BPA-PO), will be reported. For comparison purposes, the miscibility of PMMA in PEO and PPO oligomers is also analyzed. Cloud-point curves of the different blends were obtained in the composition range where phase separation was not restricted by vitrification of the homogeneous solution. In the presence of a polyisocyanate, these formulations may constitute the basis of PMMA/polyurethane blends.

Apart from their interest for application purposes, solutions of PMMA in oligodiols of different molar mass provide a particular complexity to the thermodynamic analysis due to the self-association of solvent molecules by H-bonds between terminal OH groups and ether groups present along the structure. The influence of the solvent self-association on miscibility can be analyzed by a variety of thermodynamic models that account for the inaccuracy of the random-mixing approximation [9-12]. In our case, experimental cloud-point curves were fitted with the Flory-Huggins (FH) model with an excess free energy accounted by an interaction parameter defined as a function of both composition and temperature.

The free energy per unit volume, $\Delta G$, of a mixture of a monodisperse solvent (component 1) and the polydisperse PMMA (component 2), may be written as:

$$(V_r/RT) \Delta G = (\phi_1/r_1)\ln\phi_1 + \Sigma(\phi_{2i}/r_{2i})\ln\phi_{2i} + g(T, \phi_2)\phi_1\phi_2 \qquad (1)$$

where $R$ is the gas constant, $T$ is temperature, $V_r = 84.7$ cm$^3$/mol, is the arbitrary reference volume (selected as the molar volume of the constitutional repeating unit of PMMA), $\phi$ is the volume fraction (1: solvent, 2: PMMA), and $r$ represents the ratio of the molar volume of a particular component or an $i$-mer, with respect to the reference volume. The first two terms in eq. (1) represent the combinatorial part of $\Delta G$. The effect of molecular weight on the interaction parameter is typically small [13], and is therefore neglected in eq. (1).

The function $g(T, \phi_2)$ is related to an interaction parameter $\chi(T, \phi_2)$ defined in terms of chemical potentials, by [13,14]:

$$\chi(T, \phi_2) = g(T, \phi_2) - \phi_1 g'(T, \phi_2) \qquad (2)$$

where the prime denotes the first derivative with respect to the volume fraction of component 2.

Several functions including different numbers of adjustable parameters have been proposed for $g(T, \phi_2)$ or $\chi(T, \phi_2)$ [9, 11, 13-15]. For the fitting of cloud-point curves of PMMA solutions in oligodiols, we found that the following function proposed by Prausnitz and co-workers [15], was suitable:

$$\chi(T, \phi_2) = (a + b/T)/(1 - c\phi_2) \qquad (3)$$

The term $a/(1 - c\phi_2)$ is related to the excess entropy contribution to the free energy while the term $(b/T)/(1 - c\phi_2)$ is related to the enthalpy contribution to the free energy. The random-mixing approximation, accounted by the combinatorial terms of eq. (1) (a negative contribution to the free energy), fails by the self-association of solvent molecules. This is corrected in the thermodynamic model by the excess entropy contribution (a positive contribution to the free energy). An increase in the value of the parameter $a$ will be therefore related to an increase in the self-association of solvent molecules. The meaning of parameter $b$ is the usual one. Higher values of $b$ accounts for a larger mismatching of solubility parameters of both components. The way in which values of *a* and *b* are affected by the chemical structure and molar mass of the different oligodiols, will be discussed.

**2. Experimental**

*2.1 Materials*

In order to analyze formulations useful for application purposes, a polydisperse PMMA was selected for this study. Its molar mass distribution was determined by size exclusion chromatography (SEC) using a universal calibration curve. The following molar mass averages (g/mol), were obtained: $M_n$ = 80200, $M_w$ = 133400, $M_z$ = 195000. The fraction of isotactic/heterotactic/syndiotactic triads was 10/36/54, as determined by $^1$H NMR. Its glass transition temperature was $T_g$ = 115 °C, defined as the midpoint of the change in specific heat ($\Delta c_p$ = 0.25 ± 0.02 J K$^{-1}$ mol$^{-1}$), determined by differential scanning calorimetry (DSC). Its mass density was $\rho_{PMMA}$ = 1.18 g/cm$^3$.

Chemical structures of different oligodiols used in this study are shown in Figure 1. These structures, including the nature of the core of PEO and PPO oligomers, were confirmed by $^1$H NMR and MALDI-TOF mass spectra. In mass spectra the only recorded peaks were those of *n*-mers with the chemical structures shown in Figure 1. With $^1$H NMR it was possible to determine that 90 mol % of PPO and BPA-PO chain ends were secondary hydroxyl groups and 10 mol% primary hydroxyl groups.

Selected BPA-based oligodiols were supplied by Seppic S.A. (Paris, France) (Table 1). They did not exhibit any crystallization peak that could interfere with the determination of cloud-point curves. The average (*x*+*y*) value was determined by chemical titration of OH end-groups.

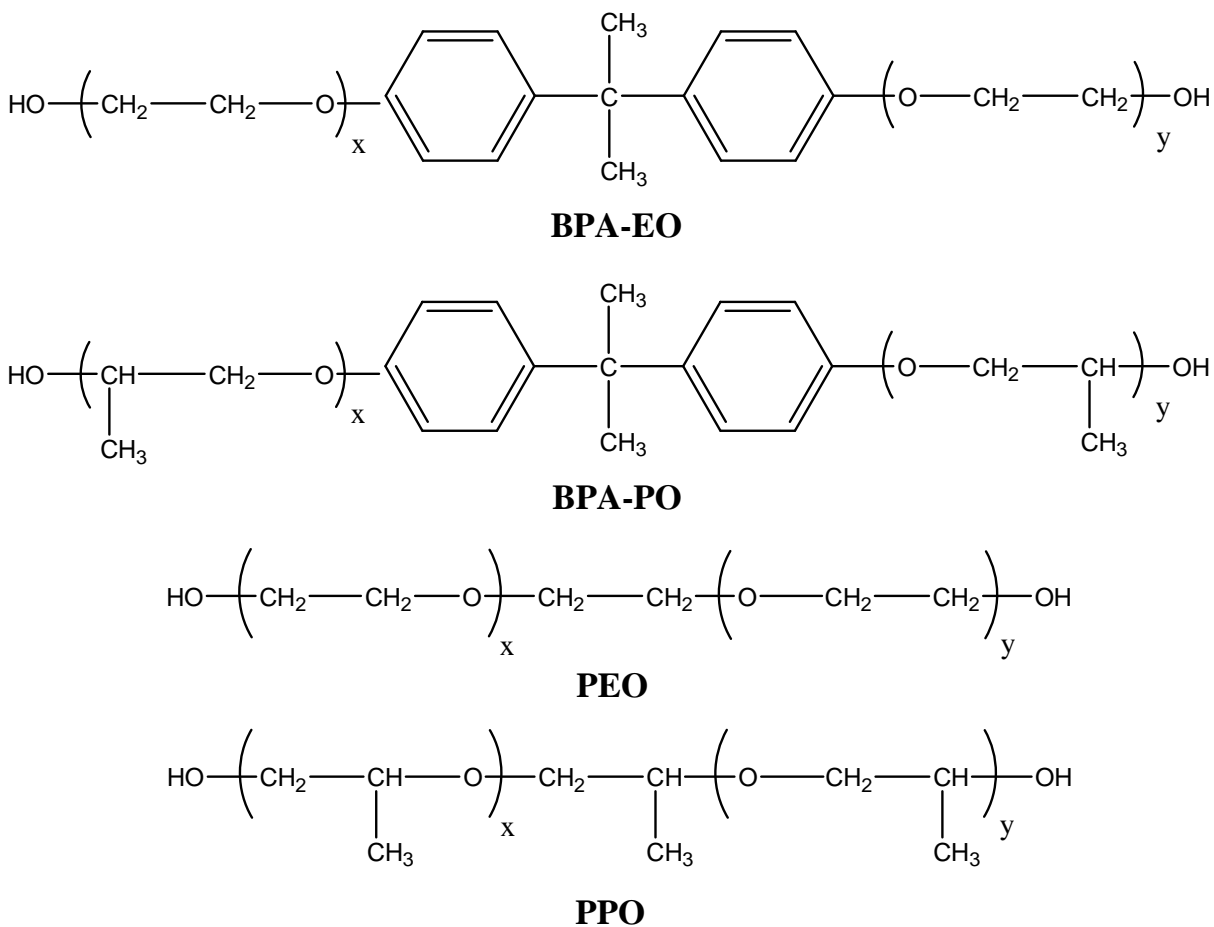

Fig.1 Chemical structures of selected oligodiols.

Poly(ethylene oxide) (PEO) and poly(propylene oxide) (PPO) oligomers, were also selected for comparison purposes (Table 2). Diethylene glycol, DEG was supplied by Carlo Erba and PEO and PPO by Aldrich. Only PEO-400 showed a melting temperature close to 0 ºC that did not interfere with the determination of cloud-point curves.

*2.2 Preparation of blends*

Blends containing less than 30 wt % PMMA were prepared in bulk while those containing higher PMMA mass fractions were obtained using 80 wt % $CHCl_3$ to aid the mixing process. The solvent was eliminated during 4 days at atmospheric pressure and one night under vacuum, at room temperature, followed by heating at 180 ºC and atmospheric pressure, until a constant weight was attained.

Table 1

Selected BPA-PEO and PBA-PPO oligodiols

| Product | $(x+y)$ (av.) | $M_n$ (g/mol) | $\rho$ (g/cm$^3$) | $T_g$ (°C) | $\Delta c_p$ (J K$^{-1}$ mol$^{-1}$) |
|---|---|---|---|---|---|
| BPA-E02 | 3.9 | 400 | 1.17 | -29 | 0.57 ± 0.03 |
| BPA-E03 | 5.9 | 488 | 1.17 | -41 | 0.60 ± 0.04 |
| BPA-E05 | 10 | 667 | 1.16 | -47 | 0.61 ± 0.06 |
| BPA-E06 | 11.7 | 741 | 1.16 | -51 | 0.71 ± 0.06 |
| BPA-PO2 | 4.1 | 465 | 1.07 | -29 | 0.56 ± 0.02 |
| BPA-PO3.5 | 6.8 | 625 | 1.06 | -44 | 0.55 ± 0.04 |
| BPA-PO5.5 | 11.1 | 870 | 1.05 | -53 | 0.55 ± 0.04 |
| BPA-PO7.5 | 15.2 | 1111 | 1.03 | -59 | 0.57 ± 0.02 |

Table 2

PEO and PPO oligomers

| Product | $(x+y)$ (av.) | $M_n$ (g/mol) | $\rho$ (g/cm$^3$) | $T_g$ (°C) | $\Delta c_p$ (J K$^{-1}$ mol$^{-1}$) |
|---|---|---|---|---|---|
| DEG | 1 | 106 | 1.12 | -98 | 0.96 ± 0.05 |
| PEO-400 | 7.7 | 400 | 1.13 | -77 | 1.00 ± 0.04 |
| PPO-400 | 6.2 | 435 | 1.00 | -74 | 0.89 ± 0.04 |
| PPO-4000 | 72.1 | 4255 | 1.00 | -77 | 0.67 ± 0.04 |

*2.3 Cloud-point curves*

Cloud-point curves were determined using a light transmission device described elsewhere [16]. Temperature was increased until a homogeneous solution was obtained, kept constant for several minutes and decreased at a rate of 1 K min$^{-1}$. The cloud point was defined at the onset temperature of the decrease in the intensity of transmitted light.

*2.4 Differential scanning calorimetry (DSC)*

Glass transition temperatures, $T_g$, defined at the mid-point of the change in specific heat ($\Delta c_p$), and melting temperatures, were determined using a Mettler DSC-30 device. Sealed aluminum pans containing 10-20 mg of sample were heated at a rate of 10 K min$^{-1}$, from –130 ºC to 50 ºC, under a continuous flow of U-grade argon. Values of $T_g$ and $\Delta c_p$ of PEO-400 were determined after quenching the sample in liquid nitrogen to avoid crystallization.

*2.5 Size exclusion chromatography (SEC)*

SEC chromatograms of PMMA were obtained with a Waters device provided with two detectors: a differential refractometer (Viscotek VE3580) and a viscometer (Dual T60); THF flow at 1 ml min$^{-1}$, $T$ = 35 ºC. The molar mass distributions was obtained using the universal calibration curve for the selected set of columns.

*2.6 Titration of OH groups*

Hydroxyl values were determined using the NF T52-112 standard. Hydroxyl groups were transformed into esters at 115 ºC, using an excess phthalic anhydride solution. The excess anhydride was determined by titration with a NaOH solution using phenolphtalein as indicator.

*2.7 $^1$H NMR*

$^1$H NMR spectra were obtained with a Bruker DRX 400 device, employing a 5mm QNP probe at 298 K. Deuterated chloroform was used as a solvent of the different oligodiols and deuterated dimethylsulfoxide was used as a solvent of PMMA.

*2.8 Matrix-assisted laser desorption ionization time-of-flight mass spectrometry (MALDI-TOF MS)*

MALDI-TOF mass spectra were obtained with a Voyager DE-STR instrument (Applied Biosystems), operating in the reflectron mode with a 337-nm laser and an accelerating voltage of 20 kV. The matrix was α-cyano-4-hydroxycinnamic acid (CHCA). A 10 g/L chloroform solution of the matrix (10 parts) was mixed with a 10 g/L chloroform solution of the analyte (1 part), and a 10 g/L acetone solution of NaI (1 part); 1 µL of the resulting solution was dropped into the target allowing solvent evaporation. Recorded spectra were the average of 300 laser shots.

**3. Results and discussion**

*3.1 Thermodynamic analysis*

Cloud-point curves were fitted with the FH model, eq. (1), considering the mixture of a monodisperse solvent characterized by the average value of $x+y$ (component 1), and the polydisperse PMMA (component 2). For PMMA, the continuous distribution of molar masses obtained by SEC was replaced by a discrete distribution composed of 36 fractions, selected in such a way that average molar masses calculated from the discrete distribution differed from values obtained with the continuous distribution in less than 1%.

The following function $g(T, \phi_2)$ was derived from eqs (2) and (3):

$$g(T, \phi_2) = (a + b/T)[1/c(1-\phi_2)] \ln[(1-c\phi_2)/(1-c)] \qquad (4)$$

Cloud-point curves were obtained equating the chemical potentials derived from eq. (1), for every component in both phases (α and β), making use of mass balances and a separation factor defined by [17,18]:

$$\sigma_2 = (1/r_{2i}) \ln(\phi_{2i}^\beta/\phi_{2i}^\alpha) \qquad (5)$$

The following two equations define the equilibrium condition:

$(1/r_1)[\ln[(1 - \phi_2 S)/(1 - \phi_2)] - \phi_2(1 – S)] + \phi_2(U – 1/r_{2n}) + (a+b/T)[(\phi_2 S)^2/(1 - c\phi_2 S)$

$- \phi_2^2/(1 - c\phi_2)] = 0$  (6)

$\sigma_2 – (1/r_1)\phi_2(1 – S) - \phi_2(U – 1/r_{2n}) + [(a+b/T)/c]\ln[(1 - c\phi_2 S)/(1 - c\phi_2)]$

$- (a+b/T)[(\phi_2 S(1 - \phi_2 S)/(1 - c\phi_2 S) - (1 - \phi_2)/(1 - c\phi_2)] = 0$  (7)

where $S = \Sigma\phi_{2i}\exp(\sigma_2 r_{2i})$, $U = \Sigma(\phi_{2i}/r_{2i})\exp(\sigma_2 r_{2i})$ and $r_{2n}$ is the ratio of the molar volume of a PMMA $i$-mer with a mass equal to the number average molar mass, with respect to the reference volume.

An experimental cloud-point curve consists of $n$ values of $T_{cp}(exp)$ determined at $n$ values of mass fraction $w_2(exp)$, converted into $\phi_2(exp)$. A single set of $a$, $b$ and $c$ values was searched for every blend, that minimized $\Sigma[T_{cp}(predicted) – T_{cp}(exp)]^2$. The system was solved using the Levenberg-Marquardt algorithm included in Mathcad 2001 Professional (predicted values of $T$ and $\sigma_2$ were obtained for each experimental composition). Standard deviations of the optimized parameters were estimated on the basis of the theory of error propagation [19-21].

Predicted cloud-point curves are represented in the corresponding Figures together with the experimental points in $T$ vs. $w_2$ coordinates.

*3.2 Glass transition temperature vs. composition*

Glass transition temperatures of homogeneous solutions that vitrified upon cooling without phase separation, were represented as a function of the initial composition using Couchman's equation [22]:

$\ln T_g = [(1 – w_2)\Delta c_{p1}\ln T_{g1} + w_2\Delta c_{p2}\ln T_{g2}]/[(1 – w_2)\Delta c_{p1} + w_2\Delta c_{p2}]$  (8)

*3.3 Cloud-point curves of different blends*

Blends of PMMA with the different oligodiols were either completely miscible or exhibited an upper-critical-solution-temperature (UCST) behavior. Figure 2 shows experimental cloud-point curves for blends of PMMA with DEG, PEO-400 and BPA-E02, together with the fitting provided by the thermodynamic model and the portion of the $T_g$ vs. $w_2$ curve represented by Couchman's equation. The selected thermodynamic model provided a reasonable fitting of cloud-point curves.

Table 3 shows the optimum set of *a*, *b* and *c* values for the different blends, given with three significant figures, together with their standard deviations. The fraction of the enthalpic contribution to the interaction parameter, defined as the ratio $(b/T)/(a+b/T)$, averaged over all experimental points, and the coordinates of the intersections of cloud-point and vitrification curves (Berghmans' points), are also indicated.

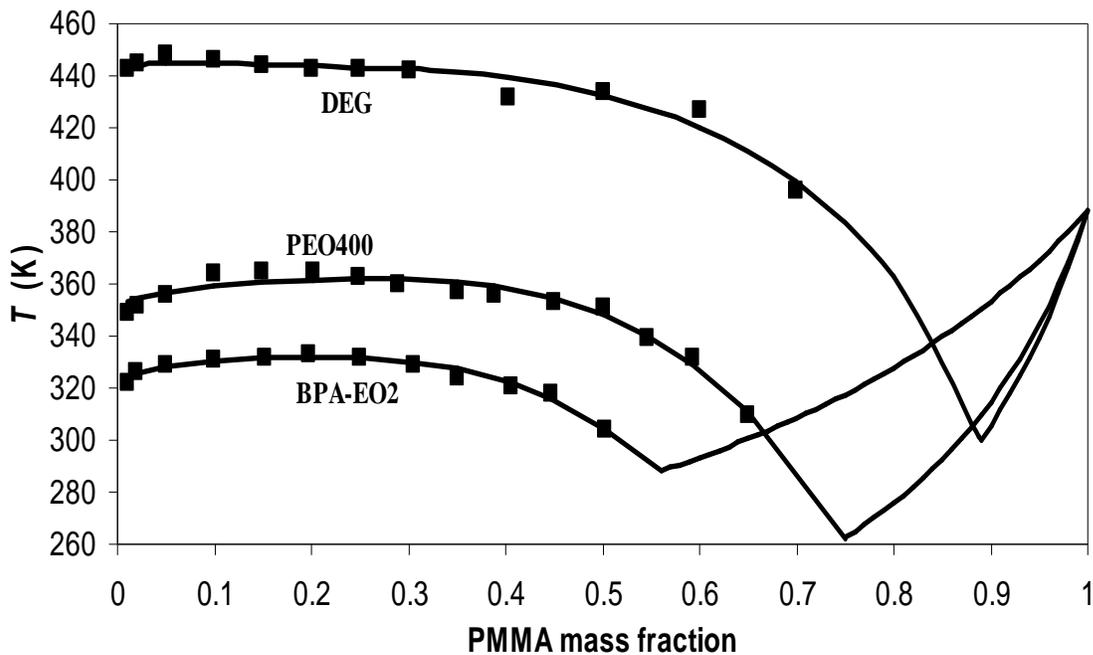

Fig.2 Cloud-point and vitrification curves for blends of PMMA with DEG, PEO-400 and BPA-EO2.

In spite of its low molar mass that favors the combinatorial entropy contribution, DEG is a bad solvent for PMMA. This is due to the high values of both *a* and *b* (estimated, in this case, with high standard deviations). As discussed in the Introduction, the relatively high value of *a* accounts for the self-association of DEG molecules through H-bonds. This parameter provides a correction for the assumption of random mixing included in the combinatorial terms. The high value of $b$ is the result of the mismatch of solubility parameters of DEG and PMMA, arising in a significant part from the contribution of terminal OH groups of DEG. A positive value of $c$ indicates an increase of the interaction parameter when increasing the PMMA concentration in the blend.

Table 3

Optimum set of $a$, $b$ and $c$ values with standard deviations, arising from the fitting of cloud-point curves of solutions of PMMA in different oligodiols; $(b/T)/(a+b/T)$ represents the average fraction of the enthalpic contribution to the interaction parameter and BP is the Berghmans Point (intersection between cloud-point and vitrification curves)

| Diol | $a$ | $b$ (K) | $c$ | $(b/T)/(a+b/T)$ | BP $T$(K) | $w_2$ |
|---|---|---|---|---|---|---|
| DEG | $0.171 \pm 0.088$ | $123 \pm 40$ | $0.680 \pm 0.018$ | 0.623 | 300 | 0.89 |
| PEO-400 | $0.0988 \pm 0.0021$ | $7.51 \pm 0.81$ | $0.707 \pm 0.003$ | 0.176 | 262 | 0.75 |
| BPA-E02 | $0.110 \pm 0.002$ | $4.87 \pm 0.50$ | $0.698 \pm 0.002$ | 0.120 | 288 | 0.56 |
| BPA-E03 | $0.0890 \pm 0.0018$ | $4.37 \pm 0.63$ | $0.699 \pm 0.003$ | 0.135 | 282 | 0.60 |
| BPA-E05 | $0.0598 \pm 0.0030$ | $4.79 \pm 1.02$ | $0.698 \pm 0.005$ | 0.181 | 284 | 0.64 |
| BPA-E06 | $0.0524 \pm 0.0037$ | $4.94 \pm 1.29$ | $0.691 \pm 0.007$ | 0.224 | 278 | 0.66 |
| PPO-400 | $0.0886 \pm 0.0010$ | $3.33 \pm 0.37$ | $0.679 \pm 0.003$ | 0.106 | 247 | 0.63 |
| PPO-4000 | $0.0085 \pm 0.0006$ | $0.98 \pm 0.23$ | $0.579 \pm 0.020$ | 0.405 | 255 | 0.63 |
| BPA-PO5.5 | $0.0453 \pm 0.0004$ | $2.10 \pm 0.12$ | $0.674 \pm 0.001$ | 0.137 | 254 | 0.50 |
| BPA-PO7.5 | $0.0360 \pm 0.0011$ | $1.71 \pm 0.44$ | $0.659 \pm 0.010$ | 0.122 | 271 | 0.60 |

Increasing the average size of PEO from the dimer (DEG) to the nonamer (PEO-400), produced a high increase of the PMMA miscibility, in spite of the lower contribution (in absolute value) of combinatorial terms. A large reduction in the value of *b* was observed, that is explained by the decrease in the contribution of terminal OH groups to the solubility parameter of PEO, leading to a better matching with the solubility parameter of PMMA. Now, the entropic part of the interaction parameter accounts for about 82 % of its overall value. This implies that it is mainly the self-association of solvent molecules that limits the miscibility of PMMA with PEO-400.

It should be expected that increasing the size of PEO chains would produce a further decrease of the interaction parameter due to a decrease of the influence of terminal OH groups on both self-association (reflected in a lower *a* value), and in the contribution to the solubility parameter (reflected in a lower *b* value). PMMA/PEO blends have been extensively investigated as they are considered one of the few examples of miscible polymer blends above the melting temperature of PEO ($T_m$ close to 338 K) [23,24]. A single glass transition temperature was reported for blends rapidly cooled into the glassy state to avoid PEO crystallization [25-27]. Investigations of melt dynamics using techniques such as NMR [28-31] and rheometry [32], indicated extensive molecular mixing. Small negative values of the interaction parameter were predicted using equation-of-state theories [33-36]. The value of the interaction parameter was experimentally obtained by small-angle neutron-scattering (SANS) measurements. Very small values were reported (either negative or positive depending on PMMA concentration), that increased with the PMMA concentration and that were dominated by the entropic term [37]. The last two findings agree with the results obtained for the PMMA/PEO-400 blend.

BPA-E02 was a better solvent for PMMA than PEO-400. Factors improving the miscibility are the slight reduction in molar volume (354 $cm^3$/mol for PEO-400 compared with 342 cm$^3$/mol for BPA-E02), and the decrease in the value of $b$, observed when EO units are replaced by the bisphenol A structure (Table 3).

Cloud-point and vitrification curves for blends of PMMA with BPA-E03, BPA-E05 and BPA-E06 are shown, respectively, in Figures 3, 4 and 5 (the representation in separate Figures is due to the overlapping of experimental points). Increasing the number of EO units attached to the bisphenol A core from 3.9 to 11.7, did not produce any significant effect on the miscibility with PMMA. The value of $b$ was practically the same for the four BPA-EO solvents (Table 3). When increasing the solvent size, there seems to be a compensation effect between the decrease of the absolute value of the combinatorial contribution to miscibility and the decrease in the self-association of solvent molecules reflected by the decrease of $a$ values (Table 3). This leads to practically the same location of cloud-point curves.

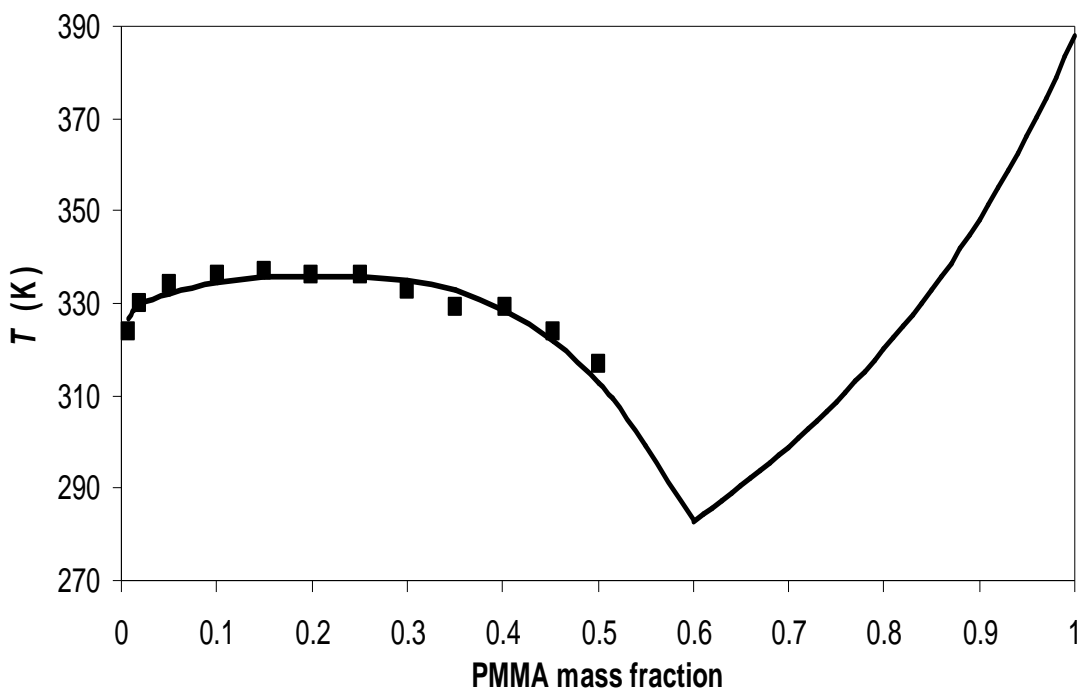

Fig.3 Cloud-point and vitrification curve for the blend of PMMA with BPA-EO3.

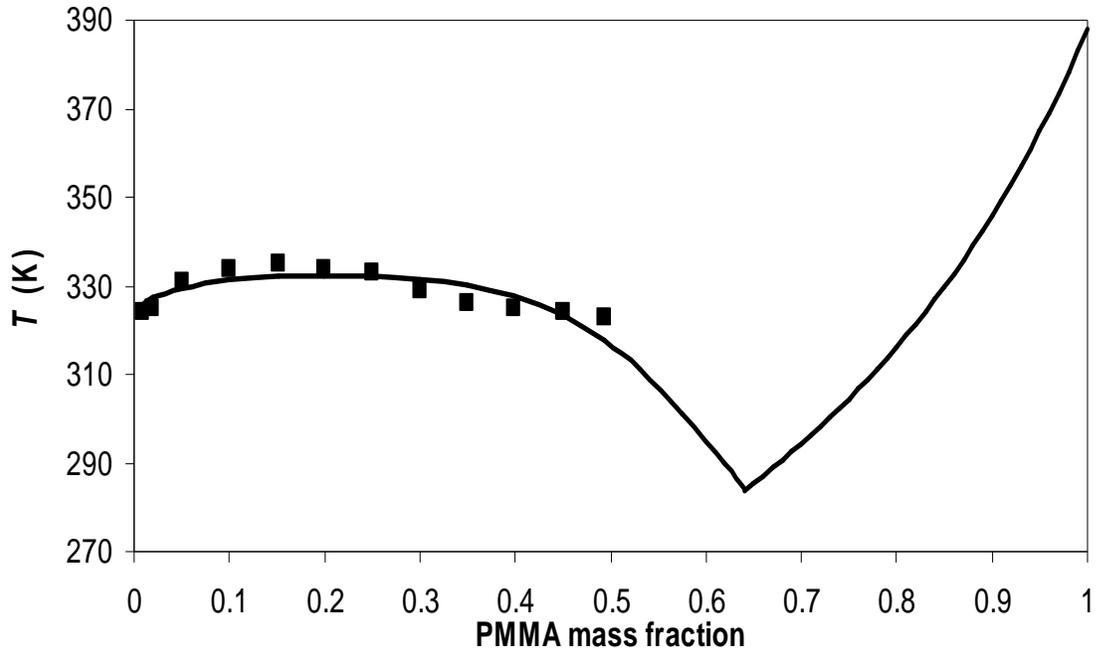

Fig.4 Cloud-point and vitrification curve for the blend of PMMA with BPA-EO5.

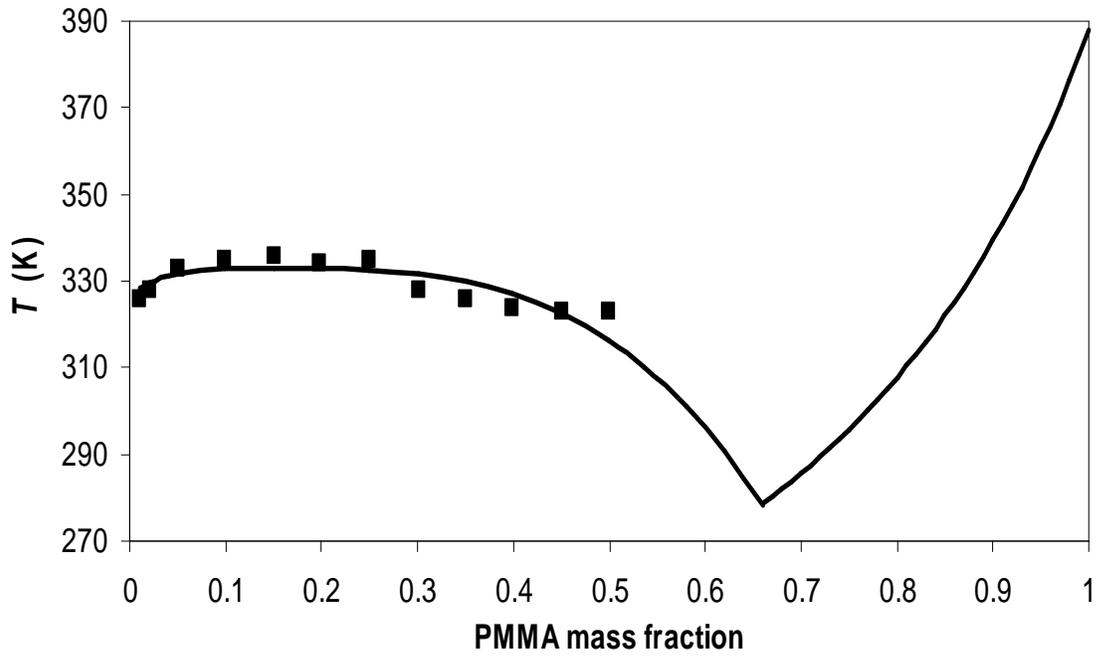

Fig.5 Cloud-point and vitrification curve for the blend of PMMA with BPA-EO6.

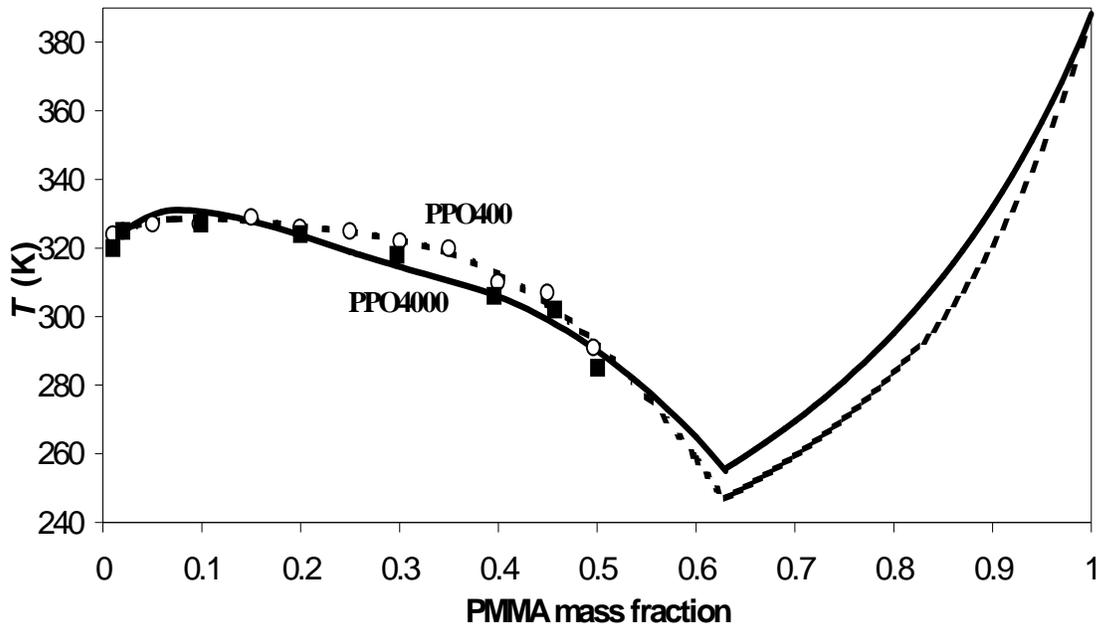

Fig.6 Cloud-point and vitrification curves for PMMA/PPO blends.

Figure 6 shows cloud-point and vitrification curves for PMMA/PPO blends. Cloud-point curves are located in the same temperature range in spite of the fact that PPO-4000 is 10 times larger than PPO-400. As shown in Table 3, this is due to the significant decrease in both *a* and *b* when increasing the solvent size, ascribed to the dilution of terminal OH groups. This effect compensates the lower contribution of combinatorial terms to miscibility.

Cloud-point and vitrification curves for the series of PMMA/BPA-PO blends are shown in Figure 7. BPA-PO solvents exhibited a higher compatibility with PMMA than the family of BPA-EO solvents. No phase separation was observed for the two solvents with lower sizes. Vitrification curves that were experimentally obtained in the whole composition range, could be fitted with Couchman's equation. Increasing the size of PPO branches produced a decrease in the PMMA miscibility. In this case, miscibility is significantly affected by the effect of solvent size on combinatorial terms. In every case, the value of the interaction parameter was dominated by the entropic contribution.

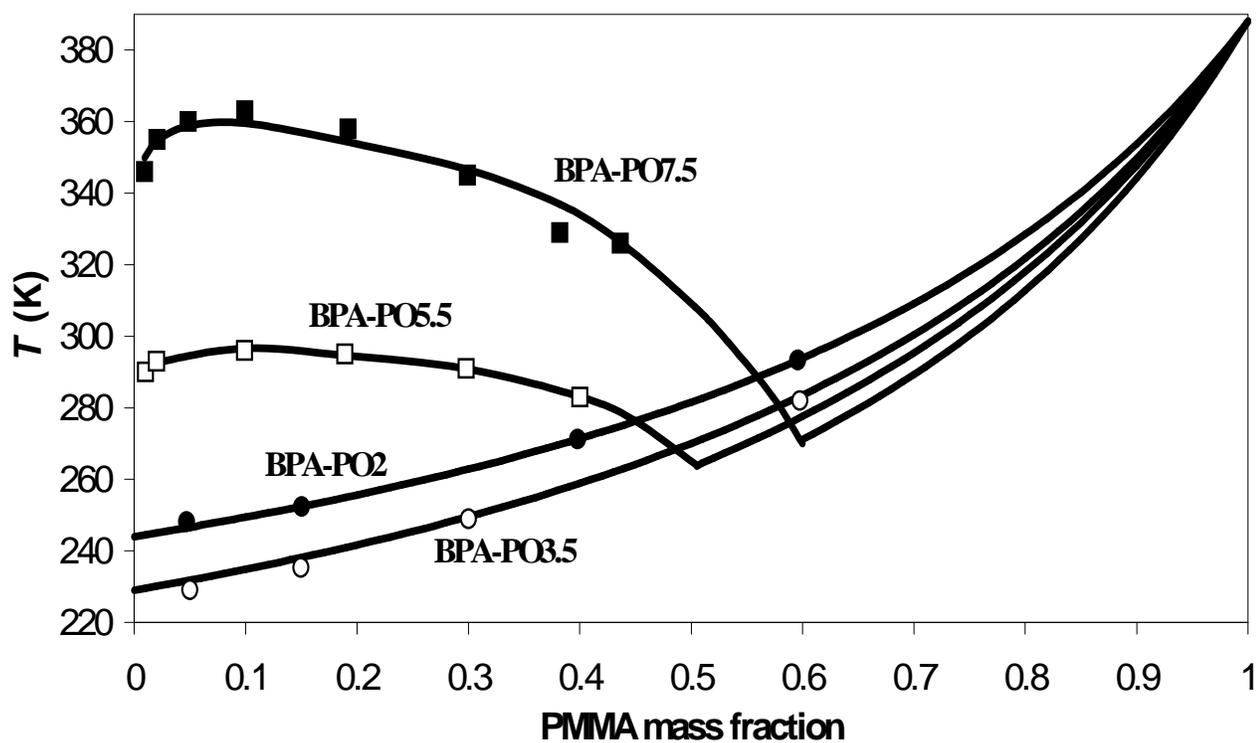

Fig.7 Cloud-point and vitrification curves for PMMA/BPA-PO blends.

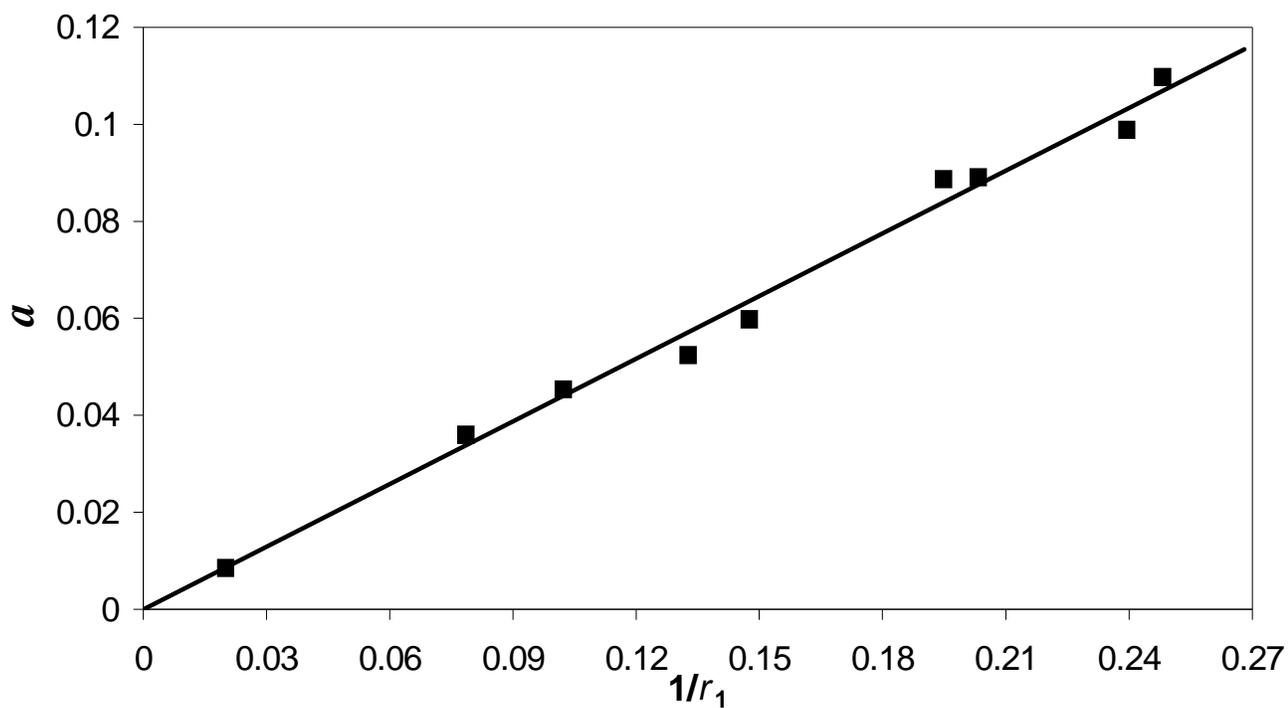

Fig.8 Variation of the entropic contribution to the interaction parameter ($a$) with respect to the inverse of the solvent size ($1/r_1$).

An inspection of the whole set of $a$ values given in Table 3 shows a clear trend of variation with the size of the solvent. Figure 8 shows a linear relationship of $a$ values for the different oligodiols with the inverse of their size ($1/r_1$). The value corresponding to DEG was not represented due to its high standard deviation. When $r_1 \to \infty$, $a \to 0$, due to the negligible self-association of solvent molecules in the absence of terminal hydroxyl groups. Experimental points for EO (primary hydroxyl groups) or PO (secondary hydroxyl groups) based oligodiols lie on the same line. This means that the entropic contribution to the interaction parameter, accounting for the solvent self-association, is independent of the chemical structure of the different oligodiols and only related to their sizes.

## 4. Conclusions

A series of cloud-point curves obtained for PMMA solutions in different oligodiols, could be reasonably fitted using the Flory-Huggins model with an interaction parameter depending on both temperature and composition. For PMMA/PEO and PMMA/PPO blends, the miscibility increased when increasing the size of the diol, due to the significant decrease in the entropic and enthalpic terms contributing to the interaction parameter. This reflected the decrease in the self-association of solvent molecules and in the contribution of terminal OH groups to the mismatching of solubility parameters. For PMMA/BPA-EO blends, a decrease of the entropic contribution to the interaction parameter when increasing the size of the oligodiol was also found. However, the effect was counterbalanced by the opposite contribution of combinatorial terms leading to cloud-point curves located in approximately the same temperature range. For PMMA/BPA-PO blends, the interaction parameter exhibited a very low value. In this case, the effect of solvent size was much more important on combinatorial terms than on the interaction parameter, leading to an increase in miscibility when decreasing the oligodiol size. For short BPA-PO oligodiols no phase separation was observed. The entropic term of the interaction parameter exhibited an inverse relationship with the size of the oligodiols, independent of the nature of the chains bearing the hydroxyls and the type of OH groups (primary or secondary). This indicates that the degree of self-association of solvent molecules through their OH terminal groups, was mainly determined by their relative sizes.

**Acknowledgements:** We are grateful to Seppic S.A. (Paris, France) for the samples of BPA-EO and BPA-PPO, to F. Boisson and M. F. Llauro (Service RMN de la Fédération de Recherche des Polyméristes Lyonnais, FR-CNRS 2151, Vernaison, France) for $^1$H NMR measurements, and to F. Delolme (Service Central d'Analyse, Vernaison, France) for MALDI-TOF MS characterization. The financial support of CONICET and ANPCyT (Argentina), and CNRS (France), is gratefully acknowledged.